\def\lncs{0}

\ifnum\lncs=1
\documentclass{llncs}
\else
\documentclass[11pt,letterpaper]{article}
\fi

\usepackage{latexsym,amssymb,epsfig,amsmath}

\ifnum\lncs=0
\usepackage[colorlinks=true,breaklinks=true,linkcolor=black, citecolor=black,filecolor=black,menucolor=black, pagecolor=black,urlcolor=black]{hyperref}
\usepackage[hyperpageref]{backref}
\fi

\ifnum\lncs=0
\advance \topmargin by -\headheight
\advance \topmargin by -\headsep
\evensidemargin \oddsidemargin
\fi

\newcommand{\an}[1]{}

\newcommand{\be}{\begin{equation}}
\newcommand{\ee}{\end{equation}}
\newcommand{\bea}{\begin{eqnarray}}
\newcommand{\eea}{\end{eqnarray}}
\newcommand{\bq}{\begin{quotation}}
\newcommand{\eq}{\end{quotation}}

\newtheorem{conj}{Conjecture}
\newtheorem{defi}{Definition}
\newtheorem{theo}{Theorem}
\newtheorem{cor}[theo]{Corollary}

\newcommand{\bconj}{\begin{conj}\protect$\!\!${\em\bf :}$\;\;$}
\newcommand{\econj}{\end{conj}}

\ifnum\lncs=1
\newenvironment{proofof}[1]{\begin{proof}[of {#1}]}{\end{proof}}
\else
\newenvironment{proof}{\noindent \textbf{Proof}:}{$\Box$ \vskip 12pt}
\newenvironment{proofof}[1]{\vskip 3pt \noindent \textbf{Proof} (of
  {#1}):}{$\Box$ \vskip 3pt}

\fi

\newcommand{\GF}[1]{\mathbb{F}_{#1}}
\newcommand{\paren}[1]{\left( {#1} \right)}
\newcommand{\tuple}[1]{\paren{ \ {#1}\ }}
\newcommand{\set}[1]{\left\{ {#1} \right\}}
\newcommand{\floor}[1]{\left\lfloor{#1} \right\rfloor}

\newcommand{\bra}[1]{\langle #1 |}
\newcommand{\ket}[1]{| #1 \rangle}

\newcommand{\proj}[1]{\ket{#1}\bra{#1}}
\newcommand{\A}{Alice}
\newcommand{\B}{Bob}

\newcommand{\acc}{\textsc{acc}}
\newcommand{\rej}{\textsc{rej}}
\newcommand{\Tr}{{\mathrm{Tr}}}

\newcommand{\qas}{\textsc{qas}}
\newcommand{\prob}[1]{\Pr \left[ #1 \right] }

\newcommand{\eps}{\epsilon}
\newcommand{\logeps}{\log{(\frac 1 \eps)}}
\newcommand{\ignore}[1]{}
\newcommand{\from}{\leftarrow}

\newenvironment{CompactEnumerate}{
  \vspace{-4pt}
  \begin{list}{\arabic{enumi}.}{%
      \usecounter{enumi} %
      \setlength{\leftmargin}{12pt}%
      \setlength{\itemsep}{0pt}
      }}
  {\end{list}}

\newenvironment{CompactItemize}{
  \vspace{-4pt}
 \begin{list}{$\circ$}{%
      \usecounter{enumi}
      \setlength{\leftmargin}{12pt}%
      }}
  {\end{list}}

\begin{document}
\title{Approximate Quantum Error-Correcting Codes \\ and Secret Sharing Schemes}

\ifnum\lncs=1 

\author{Claude Cr\'epeau\inst{1} \thanks{Supported in part by Qu\'ebec's MDER, FQRNT, Canada's NSERC, MITACS, CIAR
and the Bell University Laboratories.
Some of this research was done while the author
was visiting MSRI, Berkeley CA.}
 \and Daniel Gottesman\inst{2} 
 \thanks{Some of this research was done while the author
was supported by the Clay Mathematics Institute,
and some while the author was visiting MSRI, Berkeley CA.}
 \and
Adam Smith\inst{3}
\thanks{Some of this research was done while the author
was a student at the MIT CSAIL.}
}
\institute{McGill University,
Montr\'eal, QC, Canada.
\email{crepeau@cs.mcgill.ca}
\and
Perimeter Institute, Waterloo, ON,
Canada. \email{dgottesman@perimeterinstitute.ca}
\and
Weizmann Institute of Science,
Rehovot, Israel. \email{adam.smith@weizmann.ac.il}
}

\else 
\author{
Claude Cr\'epeau\,%
\thanks{\, School of Computer Science, McGill University,
Montr\'eal (Qc), Canada H3A 2A7.
\,Supported in part by Qu\'ebec's MDER, FQRNT, Canada's NSERC, MITACS, CIAR
and the Bell University Laboratories.
Some of this research was done while the author
was visiting MSRI, Berkeley CA.
E-mail: {\tt crepeau@cs.mcgill.ca}.
}
,\ Daniel Gottesman\,%
\thanks{\, Perimeter Institute, Waterloo, Ontario,
Canada N2J 2W9.
\,Some of this research was done while the author
was supported by the Clay Mathematics Institute,
and some while the author was visiting MSRI, Berkeley CA.
E-mail:
{\tt dgottesman@perimeterinstitute.ca}.
}
,\ Adam Smith\,%
\thanks{\,Dept of Computer Science and Applied Mathematics,
Weizmann Institute of Science,
PO Box 26, Rehovot 76100, Israel.
Some of this research was done while the author
was a student at the MIT Computer Science and AI Lab.
E-mail: {\tt adam.smith@weizmann.ac.il}.
}
}
\fi 

\date{\today}

\maketitle
\begin{abstract}
It is a standard result in the theory of quantum error-correcting
codes that no code of length $n$ can fix more than $n/4$ arbitrary
errors, regardless of the dimension of the coding and encoded
Hilbert spaces.  However, this bound only applies to codes which
recover the message exactly.  Naively, one might expect that
correcting errors to very high fidelity would only allow small
violations of this bound.  This intuition is incorrect: in this
paper we describe quantum error-correcting codes capable of
correcting up to $\floor{(n-1)/2}$ arbitrary errors with fidelity
exponentially close to 1, at the price of increasing the size of
the registers (i.e., the coding alphabet).  This demonstrates a
sharp distinction between exact and approximate quantum error
correction. The codes have the property that any $t$ components
reveal no information about the message, and so they can also be viewed
as error-tolerant secret sharing schemes.

The construction has several interesting implications for
cryptography and quantum information theory. First, it suggests
that secret sharing is a better classical analogue to quantum
error correction than is classical error correction. Second, it
highlights an error in a purported proof that verifiable quantum
secret sharing (VQSS) is impossible when the number of cheaters
$t$ is $n/4$. In particular, the construction directly yields an
\emph{honest-dealer} VQSS scheme for $t=\floor{(n-1)/2}$. We
believe the codes could also potentially lead to improved
protocols for dishonest-dealer VQSS and secure multi-party
quantum computation.

More generally, the construction illustrates a difference between exact and approximate requirements in quantum cryptography and (yet again) the delicacy of security proofs and impossibility results in the quantum model.
\end{abstract}

\section{Introduction}

Quantum computers are likely to be highly susceptible to errors
from a variety of sources, much more so than classical computers.
Therefore, the study of quantum error correction is vital not
only to the task of quantum communications but also to building
functional quantum computers.
In addition, quantum error correction
has many applications to quantum cryptography.  For instance, there is a strong connection between quantum error-correcting codes and secret sharing schemes~\cite{CGL99}, and that connection was combined with fault-tolerant quantum
computation to perform multiparty secure quantum
computations \cite{CGS02}.  Many quantum key distribution schemes also rely on ideas from quantum error-correction for their proofs of security. Thus, bounds on the performance of quantum error-correcting codes (QECCs)
in various scenarios are relevant both to the foundations of quantum information theory and to quantum cryptography.

It is an immediate result of the no-cloning theorem \cite{WZ} that
no quantum error-correcting code of length $n$ can fix $n/2$
erasures: such a code would allow one to reconstruct two
copies of an encoded quantum state from two halves of the full
codeword, which would be cloning the state. This result is valid regardless of the dimension of
the coding Hilbert space.
Another well known result from the theory of quantum error correction is
that a length $n$ code can fix $t$ arbitrary single position errors if
and only if it can fix $2t$ erasure errors \cite{Got}. This follows immediately
from the quantum error-correction conditions \cite{Got}
\begin{equation}
\bra{\psi_i} E_a^\dagger E_b \ket{\psi_j} = C_{ab} \delta_{ij}
\end{equation} 
(for basis encoded states $\{\ket{\psi_i}\}$ and correctable errors $\{E_a\}$)
and implies that no QECC of length $n$ can fix more than $n/4$ arbitrary errors,
regardless of the dimension of the coding and encoded Hilbert spaces. In contrast, a classical repetition code can correct up to $\floor{(n-1)/2}$ errors.

In this paper, we describe QECCs of length $n$ that can
correct arbitrary errors which affect up to  $t=\floor{(n-1)/2}$ positions, with the guarantee that the fidelity of the reconstructed state will be exponentially
close to 1.  That is, {\em approximate} quantum error-correcting codes
have the capability of correcting errors in a regime where no {\em exact}
QECC will function. The scheme is also a secret-sharing scheme, in that no $t$ positions reveal any information at all about the message.
The result has a number of implications for both cryptography and quantum information theory:
\begin{CompactItemize}

\item It may be possible to build approximate QECCs which are
highly efficient and yet useful in common error correction scenarios,
improving on exact QECCs for the same scenarios.  In most cases, exact
reconstruction of the quantum state is not necessary, so a more efficient
approximate QECC would be welcome.

\item The connection between correcting general errors
and erasure errors breaks down for approximate QECCs. This
suggests there is no sensible notion of distance for an
approximate quantum error-correcting code.

\item The proof of the impossibility of {\em verifiable quantum secret sharing} (VQSS) with $t\geq n/4$ cheaters in \cite{CGS02} is incorrect, since it assumes that the $t<n/4$ bound on error correction extends to approximate quantum codes. In particular, the construction described here immediately yields an honest-dealer verifiable quantum secret sharing scheme which is secure for
$t=\floor{(n-1)/2}$.

Similar constructions may allow verifiable quantum secret sharing (VQSS) with a \emph{dishonest} dealer and secure multiparty quantum computation (MPQC) beyond previously known bounds.  We have devised candidate protocols for
these tasks allowing up to $(n-1)/2$ cheaters, but we do not
present them here, as we have not yet proved their security and
they are, in any case, quite complex.


\item Secret sharing may serve as a better classical analogue to quantum error correction than does classical error correction. The sharp difference we see between perfect and approximate quantum error correction parallels to some extent a similar difference between error-tolerant secret sharing schemes (explained below) with zero error and those with exponentially small error~\cite{RB89}.  The codes here use such secret sharing schemes as a building block.

\item More generally, our results demonstrate that there can be a dramatic difference in behavior
between the exact performance of some quantum-mechanical task and approximate
performance of the task, even when the approximation is exponentially good.
A similar divergence between exact and approximate bounds has recently been
seen in the context of private quantum channels \cite{HLWS04}.  These examples serve as a caution --- especially valid in cryptography --- that intuition about approximate performance of quantum protocols may be misleading.


\end{CompactItemize}

The idea of using a randomized encoding algorithm is not new in
QECC. In particular \cite{BBPSSW} have devised codes that can
correct more (malicious) errors on average than any deterministic
QECC. However, their model significantly differs from ours in one
of two ways: they assume either that the errors occur at random or
that the code is randomly agreed on by the coder and the decoder but
is kept secret from the adversarial noise source. This model does
not seem suitable in cryptographic applications such as VQSS and
MPQC \cite{CGS02}. In our model no secret is shared by the coder
and decoder.  However, part of our code can be viewed as providing
a way for the coder to information-theoretically encrypt the
necessary secret.  (This is possible since the adversary only has
access to part of the transmitted state, though it could be any
part.)

A closer analogue to our codes is present in~\cite{LNCY}, which
gave a pure-state encoding to approximately correct a specific
error model more efficiently than a typical minimum-distance code.
(Note, however, that the nature of the error model in fact
precludes {\em any} exact quantum error-correcting code.)  Closer
yet is~\cite{SW}, which considered approximate quantum error
correction in precisely our sense, and studied conditions for
approximate error correction to be possible.  They did not,
however, present any specific codes or suggest that approximate
QECCs might allow significant improvements in the number of
correctable registers.

%

\paragraph{Secret Sharing and Quantum Error Correction}

Classically, an $(n,d)$-secret sharing scheme splits a secret into
$n$ pieces so that no $d-1$ shares reveal any information about
the secret, but any $d$ shares allow one to reconstruct it. Such a
scheme is already an error-correcting code, since it allows one to
correct up to $n-d$ erasures. Error-correcting codes need not be
secret sharing schemes: a repetition code, for example, provides
no secrecy at all. In the quantum world, the connection is much
tighter. Cleve \emph{et al.}~\cite{CGL99} observed that any
(perfect) QECC correcting $t$ erasures is itself a secret sharing
scheme, in that no $t$ components of the code reveal any
information about the message. This follows from the principle
that information implies disturbance. Furthermore, most known
(perfect) classical secret sharing schemes (and ``ramp" schemes)
can be directly transformed into (perfect) QECC's with the related
parameters~\cite{Smi00}.

The quantum code construction described here illustrates a further
connection to classical secret sharing. An error-tolerant secret
sharing scheme (ETSS) can recover the secret even when $t$ shares
have been maliciously corrupted. Ordinary $(n,d)$-secret sharing
schemes are error-tolerant: such a scheme corrects $n-d$ erasures
and hence $t = (n-d)/2$ errors (this fact was first highlighted for Shamir secret sharing in \cite{MS81}). If we also want any $t$ shares to
reveal no information, then we get $t<d$, and thus $t<n/3$. This
is optimal for schemes with zero error probability. On the other
hand, if one allows a small probability of mistaken error
correction, then one can in fact get error-tolerant secret sharing
schemes which correct $t=\floor{(n-1)/2}$ errors (see the
Preliminaries for more details). Thus, the best classical analogue
for approximate quantum codes are error-tolerant classical secret
sharing schemes which correct any $t$ errors with high
probability. These have been studied more or less explicitly in
work on multi-party computation~\cite{RB89,CDDHR99,CDF01}.

It is worth noting that the construction of quantum error-tolerant secret sharing schemes has farther reaching implications than analogous classical constructions. Our approximate quantum codes
correct a number of general errors for which no exact code would
suffice, whereas the classical constructions can be better
understood as reducing the number of erasures that can be
corrected via secret sharing techniques.  A straightforward
classical repetition code already corrects up to $\floor{(n-1)/2}$ arbitrary
errors exactly, so there is no need to resort to sophisticated
techniques to achieve this with classical ECCs.

\paragraph{Results} Our construction produces quantum codes which encode $\ell$ qubits into $n$ registers of $\frac{\ell}{(n-2t)} + O(ns)$ qubits each and which correct any $t$ adversarial errors with probability $2^{-s}$ (the bound assumes $\log n<\ell<2^{s}$ for simplicity).
This is done by transforming $[[n,1,n/2]]_{n}$ QECCs on $n$-dimensional registers into better codes on $2^{O(ns)}$-dimensional registers. The codes we construct are always decodable in polynomial time, since the only necessary operations are verification of quantum authentication and erasure correction for a stabilizer code, and since erasure correction for a stabilizer code only requires solving a system of linear equations.

\section{Preliminaries}\label{sec:qa}

\paragraph{Classical Authentication}

\newcommand{\ck}{a}

For our purposes, a classical (one-time) authentication scheme is a function $h_\ck(m)$ that takes a secret key $\ck$ and a message $m$ as input (and no other randomness), and outputs a tag for the message. Typically, \A\ sends the pair $m,h_\ck(m)$ to \B, with whom she shares the key $\ck$. \B\ receives a pair $m',tag'$ and accepts the message as valid if and only if  $tag' = h_\ck(m')$. \B\ will always accept a message that really came from Alice. The scheme has error $\eps$ if, given a valid pair $m,h_\ck(m)$, no adversary Oscar can forge a tag for a different message $m'$ with probability better than $\eps$. That is, for all messages $m$ and all (computationally-unbounded, randomized) algorithms $O()$, if $\ck$ is chosen randomly from a set of keys ${\cal K}$, then: $$\Pr_{\ck\from {\cal K}}[m',tag' \from O(m,h_\ck(m))\ :\  tag' = h_\ck(m') ] \leq \eps.$$
We make no assumptions on the running time of the adversary. If the message is $\ell$ bits long, then one can find a polynomial time authentication scheme where both the key and the tags have length $O(\log\ell + \logeps)$ (see, e.g., \cite{GN}).

\medskip

For the remainder of this paper, we assume the reader is familiar with
the basic notions and notation of quantum computing (see a textbook
such as \cite{NC00} if necessary).

\paragraph{Quantum Authentication}
Intuitively, a quantum authentication scheme \cite{BCGST02} is a keyed
system which allows \A\ to send a state $\rho$ to \B\ with a
guarantee: if \B\ accepts the received state as ``valid'', the fidelity
of that state to $\rho$ is almost 1.  Moreover, if the
adversary makes no changes, \B\ always accepts and the fidelity
is exactly 1. The following definition is from Barnum et al. \cite{BCGST02}.
We first define what
constitutes a quantum authentication scheme, and then give a
definition of security.

\begin{defi}[\cite{BCGST02}]\label{def:qas}
  A \emph{quantum authentication scheme} (\qas) is a pair of
  polynomial time quantum algorithms $A$ and $V$ together with a set
  of \emph{classical} keys ${\cal K}$ such that:
 \begin{CompactItemize}
  \item
    $A$ takes as input an $m$-qubit message system $M$ and a key
    $k\in {\cal K}$ and outputs a transmitted system $C$ of $m+t$
    qubits.
  \item $V$ takes as input the (possibly altered) transmitted system
    $\hat{C}$ and a classical key $k\in {\cal K}$ and outputs two systems: a
    $m$-qubit message state $\hat{M}$, and a single (verdict) qubit $V$ which
    indicates acceptance or rejection. The classical basis states of
    $V$ are called $\ket{\acc},\ket{\rej}$ by convention.
  \end{CompactItemize}
  For any fixed key $k$, we denote the corresponding super-operators
  by $A_k$ and $V_k$.
\end{defi}

\B\ may  measure the qubit $V$ to see whether or
not the transmission was accepted or rejected. Nonetheless, we think
of $V$ as a qubit rather than a classical bit since it will allow us
to describe the joint state of the two systems $\hat{M},V$ with a density
matrix.
%
%
Given a pure state $\ket{\psi}\in {\cal H}_M$, consider the following
test on the joint system $\hat{M},V$: output a 1 if the first $m$ qubits are
in state $\ket{\psi}$ \emph{or} if the last qubit is in state
$\ket{\rej}$ (otherwise, output a 0). The projectors corresponding to this
measurement are
\begin{eqnarray*}
P_1^{\ket{\psi}} &=& \ket{\psi}\bra{\psi}\otimes \proj{\acc}\ +\ I_{\hat{M}} \otimes
\proj{\rej} \\
P_0^{\ket{\psi}} &=& (I_{\hat{M}}-\proj{\psi})\otimes (\proj{\acc})
\end{eqnarray*}
We want that for all possible input states $\ket{\psi}$ and for all
possible interventions by the adversary, the expected fidelity of
V's output to the space defined by $P_1^{\ket{\psi}}$ is high. This
is captured in the following definition of security.

\begin{defi}[\cite{BCGST02}]\label{def:secure}
  A $\qas$ is secure with error $\epsilon$ for a state $\ket\psi$ if
it satisfies:

\begin{CompactItemize}
\item \emph{Completeness:} For all keys $k\in{\cal K}$: $V_k
    (A_k(\proj{\psi})) = \proj{\psi}\otimes \proj{\acc}$

\item \emph{Soundness:} For a super-operator ${\cal O}$, let
    $\rho_{\B}$ be the state output by \B\ when the adversary's
    intervention is characterized by ${\cal O}$, that is:
$      \rho_{\B} 
      = \frac{1}{|{\cal K}|}\sum_k V_k(
      {\cal O}( A_k(\ket{\psi}\bra{\psi})))$  (this is the expectation over all values of the key of the state output by \B).
    The $\qas$ has soundness error
    $\epsilon$ for $\ket\psi$ if for all super-operators ${\cal O}$,
    $$\Tr\paren{P_1^{\ket{\psi}} \rho_{\B} } \geq 1-\epsilon$$
  \end{CompactItemize}
  A $\qas$ is secure with error $\epsilon$ if it is secure with error
  $\epsilon$ for all states $\ket\psi$. We make no assumptions on the running time
  of the adversary.
\end{defi}

In order to authenticate a message of $\ell$ qubits, the authentication scheme of \cite{BCGST02} uses a (classical) key of length $2\ell + O(\logeps)$ random bits and produces a transmitted system of $\ell+O(\logeps)$ qubits.  The large part $2\ell$ of the classical key is used to the encrypt the quantum state, which is necessary for any quantum authentication scheme to be secure~\cite{BCGST02}.  In the special case where \A\ wishes to authenticate half of a maximally entangled state $\sum \ket{i} \ket{i}$, in fact only $O(\logeps)$ classical key bits are necessarily~\cite{OH,HLM}, effectively because \A's message is already a maximally mixed state, making encryption redundant.

\paragraph{Composability of Quantum Authentication}

We will need authentication protocols that have an
additional composability property: If $(A_k, V_k)$ is a $\qas$
with error $\epsilon$ for key $k$, then the concatenated protocol
\begin{equation}
\left(\bigotimes_{i=1}^n A_{k_i}, \bigotimes_{i=1}^n V_{k_i} \right)
\end{equation}
should be a $\qas$ with error $\epsilon$ for the key $(k_1,
\ldots, k_n)$, with the understanding that the concatenated
verification protocol accepts if and only if all of the tensor components accept (i.e. the verdict qubit for the concatenated scheme is the logical {\sc AND} of the individual verdict qubits).

This sort of composability holds trivially for a classical authentication scheme, although the error may increase linearly with the number of compositions. We do not know if the same is true in general for quantum authentication schemes. However, the quantum authentication schemes of \cite{BCGST02} are indeed composable, with no blow-up in the error parameter.  This follows because they are constructed from stabilizer purity testing codes (PTCs), which clearly satisfy a corresponding property (if $Q_k$ is a stabilizer PTC with error $\epsilon$,
then $\bigotimes_{i=1}^n Q_{k_i}$ is a stabilizer PTC with error $\epsilon$).

\paragraph{Classical Secret Sharing and Error Correction}

A classical $(n,d)$-secret sharing scheme \cite{Shamir79} is a cryptographic protocol allowing a {\em dealer} to share a secret $k$ into $n$ {\em shares} $(s_1, \ldots, s_n)$ with $n$ {\em share-holders} $P_{1}, \ldots ,P_{n}$ in such a way that any $d-1$ $s_i$'s contains no information about $k$ whereas any $d$ of those $s_i$'s completely define $k$. We write $(s_1, \ldots, s_n) \in_R SS_{n,d}(k)$, a random instantiation of a set of shares for secret $k$. The original construction of Shamir~\cite{Shamir79}, based on Reed-Solomon codes, allows one to share an $\ell$-bit secret with shares that are each $\max\set{\ell,\log n}$ bits.

An important component in our construction is a classical secret sharing scheme which allows the honest players to reconstruct the secret even if the cheaters alter their shares. Specifically, consider the following game: an honest dealer takes a secret, splits it into $n$ shares $s_1,..,s_n$, and distributes the shares amongst $n$ participants over secure channels (i.e., player $i$ gets only $s_i$). Next, an adversary (adaptively) corrupts up to $t=d-1$ of the players. Finally, all players send their (possibly corrupted) shares over secure channels to a trusted arbiter who attempts to recover the secret. The secret sharing scheme is called an \emph{error-tolerant secret sharing scheme} (ETSS) and is \emph{$t$-error-correcting with error $\eps$} if the arbiter can reconstruct the correct secret with probability $1-\eps$, regardless of the adversary's strategy. In other words, an ETSS is a secret-sharing scheme which also acts as an error-correcting code correcting any $t$ errors with high probability.

Error-tolerant secret sharing has been studied under the names ``honest-dealer VSS with a non-rushing adversary''~\cite{CDF01} and ``non-interactive Las Vegas perfectly secure message transmission''~\cite{SNR04}. ``Robust secret sharing''~\cite{CPS02} is a slightly weaker variant of the problem. Another  variant, ``honest-dealer VSS with rushing'' is slightly stronger than ETSS; see \cite{CDF01} for a discussion of the differences.

A number of constructions of ETSS schemes appear in the literature.  When $t<n/3$, any ordinary secret sharing scheme is in fact an ETSS with zero error (since it is a code correcting $2t$ erasures and hence $t$ errors). This connection was first pointed out by~\cite{MS81}. When $t$ is between $n/3$ and $n/2$, one can adapt constructions from multi-party computation protocols~\cite{RB89,CDDHR99,CDF01}. We will use a simple construction for the case $t=\floor{(n-1)/2}$ from \cite{CDF01}. The dealer encodes the secret using an ordinary secret sharing scheme, and augments the shares by creating a fresh authentication key and tag for every pair of players: $P_i$ gets the key $\ck_{ij}$ and $P_j$ gets the tag $h_{\ck_{ij}}(s_j)$. If the adversary does not succesfully forge any authentication tags for keys held by honest players, then the arbiter can reconstruct the secret by accepting only shares for which at least $t+1$ of the authentication tags are valid.

The two schemes suggested above tolerate the maximum number of cheaters. On one hand, schemes with zero error can tolerate at most $n/3$ errors~\cite{RB89}. On the other hand, it is clear that no ETSS scheme can correct more than  $t=\floor{(n-1)/2}$ errors: any $n-t$ players must be able to reconstruct the secret alone (as the adversary could simply erase all its shares), and so we must have $n-t>t$. Alternatively, one can view this as an ordinary error correction bound: if the adversary could control half of the shares, he could make them all consistent with a value of his choosing (say 0) and force the arbiter to reconstruct 0.

The main complexity measure of an ETSS scheme is the share size. For a given scheme, let $CC(\ell,\eps,t)$ denote the maximum size (in bits) of a share held by any player. When $t<n/3$, the usual Shamir secret sharing scheme is a zero-error ETSS scheme with zero error and share size $CC(\ell,0,t) = \ell/(n-3t)$ (for $\ell > (n-3t)\log n$). The errors can be corrected in polynomial time since the scheme encodes data in a Reed-Solomon code.
For $t=\floor{(n-1)/2}$, the augmented scheme using authentication tags produces shares of size $CC(\ell)=\ell + O(n\logeps)$ (when $\ell>\log n$ and $\logeps > \max\set{n,\ell}$).

Based on \cite{CPS02}, Cramer et al.~\cite{CDF01} present a more compact scheme for $t=\floor{(n-1)/2}$ with share size $O(\ell+n+\logeps)$. Unfortunately, that scheme is not known to correct the errors in polynomial time. A second scheme, for $t$ further away from $n/2$, generates shares of size $CC(\ell,\eps,t)=\Omega(n\logeps+\ell/(n-2t))$. The same work \cite{CDF01} also proved a simple lower bound on the share size of ETSS schemes:  $CC(\ell,\eps,t)=\Omega(\logeps + \frac{\ell}{(n-2t)})$. This bound is tight for $\logeps > n$ and $n=2t+1$.

\section{Definition of Approximate Quantum Codes (AQECC)}

An approximate quantum error-correcting code allows \A\ to send a state $\rho$ to \B\ with the guarantee that if few enough errors occur in transmission, the fidelity of the state received by \B\ to $\rho$ will be almost 1.

Let $q=p^m$ and $Q=p^N$ for some prime $p$ and integers $m$, $N$.
We first define what constitutes an AQECC over $\GF{Q}$, and then give a
definition of correctness. (Note that the definition makes sense over any alphabet, but we restrict to prime powers for simplicity). 

\begin{defi}\label{def:aqecc}
  An \emph{approximate quantum error correcting code} (AQECC) is a pair of
  quantum algorithms $E$ (encoder) and $D$ (decoder) such that:
 \begin{CompactItemize}
  \item
    $E$ takes as input a $m$-qu$q$it message system $M$
    and outputs a (mixed state) codeword $C$ of $n$ qu$Q$its.
  \item $D$ takes as input the (possibly altered) transmitted system
    $\hat{C}$ and outputs a $m$-qu$q$it message state $\hat{M}$.
  \end{CompactItemize}
\end{defi}

In our constructions, both the encoding $E$ and error-correction algorithm $D$ run in polynomial time in the number of qubits of input.

We will define the correctness of an AQECC on pure states, but it
follows from a result of Barnum, Knill and Nielsen (\cite{BKN},
Thm 2) that the output of the AQECC also has high fidelity to an
input which is mixed or part of an entangled state.

Given a pure state $\ket{\psi}\in {\cal H}_M$, consider the following
test on the system $\hat{M}$: output a 1 if the first $k$ qu$q$its are
in state $\ket{\psi}$ (otherwise, output a 0). The projectors corresponding to this measurement are
\begin{eqnarray*}
P_\psi &=& \proj{\psi} \\
P_\psi^\perp &=& (I_{\hat{M}}-\proj{\psi})
\end{eqnarray*}
We want that for all possible input states $\ket{\psi}$ and for all
possible interventions by the adversary, the expected fidelity of
\B's output to the space defined by $P_\psi$ is high. This
is captured in the following definition of correctness.

\begin{defi}\label{def:correct}
  An AQECC is $t$-correct with error $\epsilon$ for a state $\ket\psi$ if
for all super-operators ${\cal O}$ acting on at most $t$ qu$Q$its
(that is, ${\cal O}$ can be written as $I_{n-t} \otimes {\cal
\tilde O}_t$ for some partition of the system into $n-t$ and $t$
qu$Q$its),
$$\Tr\paren{P_\psi \rho_{\B} } \geq 1-\epsilon,$$
%
where $\rho_{\B}$ is the state output by \B\ when the adversary's
intervention\footnote{We make no assumptions on the running time
of the adversary.} is characterized by ${\cal O}$, that is:
%
$$
      \rho_{\B} = D( {\cal O}(E(\proj{\psi}))).
$$

  An AQECC is $t$-correct with error $\epsilon$ if it is $t$-correct with error
  $\epsilon$ for all states $\ket\psi$.
\end{defi}

\section{A length 3 quantum code approximately correcting one arbitrary error}
\label{warmup}

We start with a small example, from a well known code.
The code $c$ corrects one erasure error:
\begin{eqnarray}
\ket{0} & \rightarrow & \ket{000} + \ket{111} + \ket{222} \nonumber \\
\ket{1} & \rightarrow & \ket{012} + \ket{120} + \ket{201} \\
\ket{2} & \rightarrow & \ket{021} + \ket{102} + \ket{210} \nonumber
\end{eqnarray}
Let $H_1 \otimes H_2 \otimes H_3$ be the coding space of the original code
$$c\ket{\psi} \in H_1 \otimes H_2 \otimes H_3,$$
and let $(A_k, V_k)$ be a quantum authentication scheme as constructed in
\cite{BCGST02}.

We construct a three-component code $c'$ as follows:
\begin{eqnarray}
c' \ket{\psi} & = & \tuple{    A_{k_1}(H_1),k_2,k_3 }, \nonumber\\
& & \tuple{    A_{k_2}(H_2),k_1,k_3 },\\
& & \tuple{    A_{k_3}(H_3),k_2,k_1 }. \nonumber
\end{eqnarray}
Let $H_1^\prime \otimes H_2^\prime \otimes H_3^\prime$ be the coding space of
the new code
$$c' |\psi\rangle \in H_1^\prime \otimes H_2^\prime \otimes H_3^\prime$$

Note that $k_1$, $k_2$, and $k_3$ are random {\em classical} strings which we
use as keys for the quantum authentication protocol $A_k$.  Thus, the
$H_i^\prime$s contain both quantum and classical information.  Intuitively,
we use the $\qas$ to ensure that an adversary cannot change the quantum state
of a single register without being detected; thus, we can transform general
errors into erasure errors, allowing us to correct one faulty register out
of three (no exact QECC can do this).  Then we distribute the authentication
keys among the three registers so that \B\ can recover them.  We must, however,
do so in a way that prevents an adversary with access to a single register
from either learning the key applying to her own register (which would allow
her to change the quantum state) or from preventing reconstruction of the
classical keys.

\begin{theo}
If $A_k$ is a $\qas$ secure with error $\epsilon$ then $c'$ is a $1$-correct
AQECC with error prob.~${\rm poly}(\epsilon)$, correcting one arbitrary error.
\end{theo}

We omit the proof of this theorem, as in Section \ref{gencase} we
will prove a more general result.

\subsection{Reconstruction}

In all cases, the reconstruction has two phases.  First we reconstruct the
classical keys and use them to verify and decode the quantum authentications.
This may result in discarding one register, but at least two remain, which is
enough for the erasure-correcting code to recover the original encoded state.
Consider the following cases:

\begin{CompactItemize}

\item All $k_i$'s agree in $H_1^\prime,H_2^\prime,H_3^\prime$:\\
Recover $k_i$ from either $H_j^\prime, j\neq i$, check that $A_{k_i}(H_i)$
properly authenticates $H_i$. If one authentication fails, ignore the improperly
authenticated $H_i$ and reconstruct the valid codeword as
$c|\psi\rangle \in H_1 \otimes H_2 \otimes H_3$ using the erasure recovery
algorithm from both $H_j, j\neq i$.

\item Some $H_i^\prime$ disagrees with $H_j^\prime,H_h^\prime$ on both keys
$k_h$ and $k_j$:\\
Discard register $i$, which must be corrupted.  Recover $k_j$ from $H_h^\prime$
and $k_h$ from $H_j^\prime$, and decode the authentications $A_{k_j} (H_j)$ and
$A_{k_h} (H_h)$ (which should both pass, since only one register can fail).
Reconstruct the valid codeword as
$c|\psi\rangle \in H_1 \otimes H_2 \otimes H_3$ using the erasure recovery
algorithm from $H_j$ and $H_h$.

\item $H_i^\prime$ and $H_j^\prime$ disagree on key $k_h$, while $H_h^\prime$
agrees with everyone:\\
Either register $i$ or $j$ is corrupt.  Get $k_i$ and $k_j$ from $H_h^\prime$
and check that $A_{k_i}(H_i)$ properly authenticates $H_i$, and that
$A_{k_j}(H_j)$ properly authenticates $H_j$. If neither fails, reconstruct the
valid codeword as $c|\psi\rangle \in H_1 \otimes H_2 \otimes H_3$ using the
erasure recovery algorithm from $H_i$ and $H_j$. If one fails, say
$A_{k_i}(H_i)$, then conclude register $i$ is corrupt and recover $k_h$ from
$H_j^\prime$, decode $A_{k_h} (H_h)$, and reconstruct the valid codeword as
$c|\psi\rangle \in H_1 \otimes H_2 \otimes H_3$ using the erasure recovery
algorithm from $H_h$ and $H_j$.

\end{CompactItemize}

Other cases cannot arise, since only one register can have been changed from
the original encoding.

\section{A general $n$-component approximate QECC family correcting up to $d-1<n/2$ arbitrary errors}
\label{gencase}

In order to generalize the above construction to cases with $n$
registers, we need to systemize the distribution of the classical
keys.  Again, it is helpful to imagine that we are trying to
defeat an adversary with access to $t<n/2$ components of the code.
Recall that we needed two conditions: First, the adversary should
not be able to learn the classical key for her register, but the
receiver \B\ should be able to reconstruct the keys. Second, the
adversary should not be able to interfere with \B's reconstruction
of the keys.

These are precisely the properties of an ETSS.  This suggests the following strategy for building a $t$-correct AQECC: encode $\ket\psi$ using a distance $t+1$ QECC, authenticate the $n$ components using keys $\vec k = k_1,...,k_n$, and then share $\vec k$ using a classical ETSS.  The result could be considered to be a quantum ETSS (that is, an ETSS for quantum data).  However, the ramifications of this construction for quantum data are more far-reaching than for the classical protocol.  Not only does the quantum ETSS have potential cryptographic applications, but it demonstrates the possibility of exceeding the no-cloning bound on QECCs.  Indeed, any QECC, exact or approximate, is in some sense a quantum ETSS --- the ability to (approximately) correct erasures on a set of registers implies that an adversary with access to those registers can gain (almost) no information about the encoded data~\cite{SW}.


Let ${\cal Q}$ be a QECC that can correct $d-1<n/2$ arbitrary erasure errors:
${\cal Q} = [[n,k,d]]$.  Such a code can be constructed over sufficiently
large dimension $Q$; for instance, use a polynomial quantum code~\cite{AB99}.
The coding space of ${\cal Q}$ is
defined as
$${\cal Q}|\psi\rangle \in H_1 \otimes H_2 \otimes H_3 \otimes ... \otimes
H_n.$$
We assume $\dim(H_1)=\dim(H_2)=...=\dim(H_n)$.

We construct a new code ${\cal Q}'$ over larger Hilbert spaces
that can correct $d-1<n/2$ arbitrary errors except with small probability.
Register $i$ of the $n$-component code ${\cal Q}'$ contains the following:
\begin{equation}
\langle A_{k_i}(H_i),s_i,[ \ck_{ij} (\forall j \neq i) ],
                [ h_{\ck_{ji}}(s_i) (\forall j \neq i) ] \rangle,
\end{equation}
where we have used the classical authentication scheme (in systematic form):
\begin{equation}
m,\ck \rightarrow (m,h_\ck (m)),
\end{equation}
which has error $\epsilon$, and
$(s_1, \ldots, s_n) \in_R SS_{n,d}(k_1,\ldots, k_n)$, a secret sharing scheme
such that any $d-1$ $s_i$'s contains no information about $(k_1, \ldots ,k_n)$
whereas any $d$ of those $s_i$'s completely define $(k_1, \ldots ,k_n)$.  The combination of classical secret sharing and classical authentication forms an ETSS~\cite{CDF01}, as described above; in fact, any ETSS would do.

For instance, the $n=3$ case of this construction is as follows:
\begin{eqnarray}
c' \ket{\psi} & = &  \tuple{    A_{k_1}(H_1),s_1,[ \ck_{12}, \ck_{13} ],
                [ h_{\ck_{21}}(s_1),  h_{\ck_{31}}(s_1) ] }, \nonumber\\
& & \tuple{    A_{k_2}(H_2),s_2,[ \ck_{21}, \ck_{23} ],
                [ h_{\ck_{12}}(s_2),  h_{\ck_{32}}(s_2) ] },\\
& &\tuple{    A_{k_3}(H_3),s_3,[ \ck_{31}, \ck_{32} ],
                [ h_{\ck_{13}}(s_3),  h_{\ck_{23}}(s_3) ] }. \nonumber
\end{eqnarray}
Note that this is more complicated than the scheme in
section~\ref{warmup}.  Instead of giving the keys $k_i$ to the
other two players, we have instead shared them among all three
players, so no single component has access to any of the three
keys used for quantum authentication.  In section~\ref{warmup}, we were able to use the fact that the quantum register attacked by the adversary must be the same as the classical register attacked, so it is only necessary to protect information about one of the keys $k_i$, not all of them.  With the extra flexibility granted the adversary by being able to attack multiple registers, it is more straightforward to protect all $n$ keys with the classical ETSS.

We are now ready for our main result.
Let $H_1^\prime \otimes H_2^\prime \otimes ... \otimes H_n^\prime$ be the
coding space of the new code
$${\cal Q}' |\psi\rangle \in H_1^\prime \otimes H_2^\prime \otimes ... \otimes H_n^\prime$$

\begin{theo}
\label{main} If $A_k$ is a $\qas$ secure with error $\epsilon$, ${\ Q}$ is a non-degenerate stabilizer code with distance $d$, and $h_a(\cdot)$ is a classical authentication scheme with error $\eps$, 
then ${\cal Q}'$ is an approximate quantum error-correcting code correcting $d-1$ arbitrary errors with error at most $2 n^2 \epsilon$.
\end{theo}

\subsection{Reconstruction}

The reconstruction procedure is similar to that for the previous
protocol, but slightly more involved, since we must verify the
classical authentications as well.  Rather than breaking the
procedure into different cases, in this version of the protocol,
we can systematically go through four steps: First, verify the
classical authentications and discard any invalid classical share.
Second, reconstruct the keys $k_i$.  Third, verify and decode the
quantum authentications.  Fourth, discard any invalid quantum
register and reconstruct the encoded quantum state.

\begin{CompactEnumerate}

\item Verify classical authentications:\\
For each $s_i$, consider it valid if at least half its
authentications are correct according to $\ck_{ji}, j\neq i$.
Discard any share $s_i$ which is not valid.

\item Reconstruct the keys $k_i$:\\
Up to $d-1$ shares $s_i$ can have been discarded in the first stage, so at least
$n-d+1 \geq n/2+1 > d$ shares remain.  Use these to reconstruct $(k_1, \ldots, k_n)$.
If the remaining shares are not all consistent with a single value of the
secret, \B\ aborts and outputs the quantum state $\ket{0}$.

\item Verify and decode the quantum authentications:\\
Use the key $k_i$ to verify and decode the quantum authentication
$A_{k_i}(H_i)$.

\item Reconstruct the encoded quantum state:\\
Discard any registers which failed the quantum authentication, and use the
remaining registers to reconstruct the valid codeword as
$c|\psi\rangle \in H_1 \otimes \ldots \otimes H_n$ using the erasure recovery
algorithm.  (At most $d-1$ have been discarded.)  If the remaining registers
are not consistent with a single quantum codeword, \B\ aborts and outputs the
quantum state $\ket{0}$.

\end{CompactEnumerate}

We prove this assuming the original QECC ${\cal Q}$ is a
nondegenerate CSS code (which is sufficient to demonstrate that
AQECCs exist correcting up to $(n-1)/2$ errors), but the proof can
easily be extended to an arbitrary stabilizer code.

\begin{proofof}{Theorem \ref{main}}
If no errors occurred, the above procedure will exactly
reconstruct the original encoded state.  We need to show that it
still approximately reconstructs the state when there are up to
$d-1$ arbitrary errors in unknown locations.  Let $B$ be the set
of registers attacked by the adversary, and let $A=[n]\setminus B$
be the registers held by honest players.

The intuition for the proof is simple. With high probability, the
authentication keys will be reconstructed correctly; conditioned
on that event, all components of the QECC which pass the
authentication test should be ``close'' to the encoding of
$\ket\psi$ restricted to those positions, and applying erasure
correction should  yield a state very close to $\ket\psi$.
Formalizing this intuition is more delicate than it would be if
the data involved were classical. The quantum version of the
statement ``such-and-such event holds with probability $1-\eps$''
is ``the state of the system has fidelity at least $1-\eps$ to the
subspace such-and-such.''  The problem lies in the fact that the
union bound from ordinary probability, which is the basis of the
intuition outlined above, does not always hold in the quantum
world. Our solution follows the lines of the ``quantum to
classical reductions'' in \cite{LC99,CGS02}. We define a set of
``target'' subspaces whose projectors commute (in other words,
there exists a single basis of the state space in which all the
projectors are diagonal), and show that the system lies close to
each of these target subspaces. For commuting subspaces, the union
bound does hold: if the system has high fidelity to each of the
subspaces, then in fact it has high fidelity to their
intersection. To complete the proof it is sufficient to show that
for states in the intersection, the initial input $\ket\psi$ is
reconstructed exactly.

The first step is to take care of the classical component of the
encoding (composed of the shares $s_i$, classical authentication
keys $\ck_{ij}$ and tags $h_{\ck_{ij}}(s_j)$). We rely on three
observations. First, we may assume w.l.o.g. that the recovery
procedure measures all the classical components in the
computational basis before doing any processing; thus, the state
received by the reconstructor \B\ is a mixture (not a
superposition) over different bit strings which he might be sent
instead of the original ones.
 Second, the classical information held by the adversary is statistically independent of $\vec k=(k_1,...,k_n)$, the vector of quantum authentication keys.  (This follows from the fact that any $t$ of the shares $s_1,...,s_n$ are independent of the shared secret.) Third, any classical authentication tags changed by the adversary will be rejected by \B\ with probability at least $1-\eps$.

We define our first target subspace $S_0$ by the statement ``the
keys $\vec k$ reconstructed by \B\ are equal to the original
keys.'' This statement can fail only if some tag changed by the
adversary is accepted by \B, and by a (classical) union bound this
can occur with probability at most $tn\eps < n^2\eps$. The
fidelity to $S_0$ is thus at least $1-n^2\eps$.

We now look at what happens within the subspace $S_0$. Consider the following set of measurements which might be performed by \B\ after verifying the authentications, but before applying erasure correction to the code. We assume for simplicity that the adversary holds the wires $B=\set{1,...,t}$, and the wires $A=\set{t+1,...,n}$ are untouched.

\newcommand{\C}{{\mathcal C}}

\begin{itemize}
    \item For each register $i\in[n]$, $\proj{\rej_i}$ measures whether or not \B\ rejected the authentication of the $i$-th quantum system (correspondingly, $\proj{\acc_i}$ measures whether or not \B\ accepts).
    \item We use the fact that the quantum error-correcting code
    is a nondegenerate CSS code.  The code can be defined by a
    sequence of parity checks performed in two bases: the standard
    computational basis and the rotated Fourier (or ``diagonal'')
    basis.  We assume there are $r$ independent parity checks in
    the rotated basis and $s$ independent parity checks in the
    standard basis.  Denote by $V$ the linear space of parity
    checks satisfied in the computational basis, and by $W$ the
    corresponding set for the Fourier basis.  If the QECC code has
    distance at least $t+1$, then there is a basis $v_1,...,v_s$
    of $V$ such that, for any $i\in B$, position $i$ is only in
    the support of $v_i$. Same for $W$: there is a basis of parity
    checks $w_1,...,w_r$ such that only $w_i$ involves the
    $i$-th component of the code for $i\in B$. We denote by
    $\Pi_{v_i}$, $\Pi_{w_i}$ the corresponding projectors (that
    is, $\Pi_{v_i}$ preserves the supspace in which the parity
    check $v_i$ is satisfied).
\end{itemize}

The sets of projectors $\set{\proj{\rej_i}}_{i\in [n]}$,
$\set{\Pi_{v_i}}_{i \in [s]}$ and $\set{\Pi_{w_i}}_{i\in[r]}$ all
commute with each other. The only possible interaction comes from
the fact that the operators $\set{\Pi_{v_i}}$ and
$\set{\Pi_{w_i}}$ operate on the same space, but they commute by
definition of CSS codes. We may ignore projectors with indices
$i>t$ since they correspond to checks which will always be passed
within the subspace $S_0$: Therefore the system will have fidelity
$1$ to the subspaces defined by $\set{\Pi_{v_i}}$ and
$\set{\Pi_{w_i}}$ for $i>t$.

We would like to claim that, whenever \B\ accepts the set $R$ of
registers, $R$ satisfies all the parity checks restricted to $R$.
We can quantify this as follows: for all $i$ between 1 and $t$,
the system should lie in the subspace defined by
\begin{equation}
P_i = (\Pi_{v_i}\Pi_{w_i} \otimes \proj{\acc_i}) + (I \otimes
\proj{\rej_i}).
\end{equation}
where $I$ is the identity operator.
The security of the quantum authentication scheme, and the fact
that the adversary doesn't learn anything about the keys from the
classical secret sharing, imply that the fidelity to each of these
subspaces is at least $1-\eps$ (note: this requires the quantum
authentication scheme to be secure even when composed up to $t$
times). For $1\leq i \leq t$, we can define the subspaces
$S_1,\ldots,S_t$ corresponding to the projectors $P_1, \ldots,
P_t$. By a union bound, the state of the whole system has fidelity
at least $1-n^2\eps - t\eps$ to the intersection $S= \bigcap
_{i=o}^t S_i$. In words, $S$ is the space of states for which Bob
reconstructs the correct authentication keys, and for which the
set of registers accepted by \B\ satisfies all the parity checks
restricted to that set.

It remains to prove that within the space $S$, \B\ will always recover the input state $\ket\psi$ exactly. We may assume w.l.o.g. that \B\ will measure all $n$ of the registers which indicate whether the authentication failed or not in the basis $\set{\ket\rej, \ket\acc}$. Thus, the global state may be seen as a mixture over possible sets of registers accepted by \B.  If \B\ also performs the measurements $P_i$, he will, with probability at least $1-n^2\eps - t\eps$, find that the state actually satisfies all parity checks restricted to the set $R$ of registers he accepts.

When this occurs, it then follows that applying erasure correction to $R$ yields the same result as if we had used only registers untouched by the adversary. For a detailed proof of this fact, we refer the reader to Proposition 2.2 in \cite{CGS02}. The intuition behind it is straightforward: Suppose $s$ registers are discarded, leaving up to $t-s$ registers attacked by the adversary.  But because $s + (t-s) < d$, the QECC can both correct $s$ erasures and detect an additional $t-s$ errors, so the adversary is unable to reach any state in $S$ except the correct input state $\ket\psi$.  We can conclude that Bob recovers a state $\rho$ with fidelity at least $1-2n^2\eps$ to $\psi$, as desired.
\end{proofof}

\subsection{Specific Constructions and Parameters}

As mentioned above, it is natural to instantiate our construction using the polynomial codes (quantum Reed-Solomon codes) of Aharonov and Ben-Or \cite{AB99}. These are nondegenerate CSS codes over an alphabet of size $q$ whenever $q$ is a prime power and greater than $n-1$.
For any $t< n/2$, one can find a $[[n,n-2t,t+1]]_q$ code (i.e. which encodes $(n-2t)\log q$ qubits and has distance $t+1$).  This means that to encode $\ell>n$ qubits, each component of the code will consist of $\ell / (n-2t)$ qubits.  The components of the approximate QECC then consist of $\ell/(n-2t) + O(\logeps)$ qubits and $CC(2\ell/(n-2t)+O(\logeps), \eps,t)$ bits (where $CC()$ is the share size of the classical ETSS).

For $2t < n-1$, we can modify the ETSS above to get shares of size $O(n\logeps)\allowbreak + \ell/ (n-2t)$. Putting these constructions together, we can get quantum codes where each register contains $O(n(\ell/(n-2t) + \logeps))$ qubits.

An immediate improvement can be made to these parameters by noting that, for any distance $d$ nondegenerate stabilizer code, including the polynomial codes used here, the state of any $d-1$ registers is maximally entangled with the remaining registers.  Therefore, as noted in section~\ref{sec:qa}, a much shorter classical key suffices for quantum authentication.  In particular, a classical key of length $O(\log\ell+\logeps)$ is sufficient to authenticate $\ell$ EPR halves. This leads to an approximate quantum code where each component consists of $\ell/(n-2t) + O(\logeps)$ qubits and $CC(n\logeps,\eps,t)$ bits (when $\eps < 1/\ell$). This gives a total size of $\ell/(n-2t) + O(n\logeps)$.

\begin{cor}[to Theorem~\ref{main}]
   For $t< n/2$, there exists an approximate QECC correcting any $t$ errors with error $\eps$, where each component consists of $O(\ell/(n-2t) + n\logeps))$ qubits. When $n=2t+1$, we get components of size $O(\ell+ n\logeps)$.
\end{cor}

\section{Discussion and open questions}

We have constructed quantum error correcting codes that are
capable of correcting general errors when up to half the registers
are affected.  This contrasts considerably with known upper bounds
that limit a QECC to correcting errors on less than one-fourth of
all registers.  The price for being able to violate this bound is
that we only correct the state approximately; however, we do so
with exponentially good fidelity.

In general, extrapolating from exact performance of a quantum task
to approximate performance is dangerous, but possible. Factors of
the dimension may arise, and since the dimension is exponential in
the number of qubits, dramatically different behavior becomes
possible.  This phenomenon is likely behind the performance of our
codes, and suggests that high-fidelity AQECCs are only possible
when working in high dimension.

\an{Used to say: ``It remains an interesting open question, however, if it is actually possible to construct AQECCs with both high efficiency and high fidelity in the usual model where the encoded subspace has the same dimension as each register of the code."\\
There are lower bounds on the alphabet size for the classical case \cite{CDF01}. I think that the same argument can be used for the quantum case.}

Our codes instead consist of a small logical subspace and large
registers containing both quantum and classical information. As
such, they are not so useful for practical problems in quantum
error correction, but do serve as an interesting in-principle
demonstration of the potential power of approximate error
correction.  In addition, they act as quantum ETSS schemes, and may be a useful stepping stone towards building VQSS and MPQC with a large number of cheaters.  Any such construction must be more complex, however, to
take account of dishonest senders and receivers, and to allow the
participants in the protocol to alter a state in the correct way
without altering it in any unapproved manner.  Indeed, it remains
possible that the prior bound of $n/4$ cheaters does in fact
restrict VQSS and MPQC; however, we have shown here that the
existing proof of that bound does not apply to VQSS and MPQC
protocols which only guarantee approximate reconstruction of the
quantum state.

\newcommand{\STOC}[2]{{\it Proc. #1 ACM Symp. on Theory of Computing}, #2}
\newcommand{\FOCS}[2]{{\it Proc. #1 IEEE Symp. on
Foundations of Computer Science}, #2}
\newcommand{\crypto}[1]{{\it Advances in Cryptology --- CRYPTO {#1}}}
\newcommand{\eurocrypt}[1]{{\it Advances in Cryptology --- EUROCRYPT {#1}}}
\newcommand{\tcc}[1]{{\it Theory of Cryptography {#1}}}

\newcommand{\soda}[1]{{\it Proc. ACM Symp. on Discrete Algorithms,
    {#1}}}
\newcommand{\ccs}[1]{{\it Proc. ACM Conf. Computer and
    Communications Security, {#1}}}

\newcommand{\infotheo}{{\em IEEE Transactions on Information Theory}}

\newcommand{\jcss}{{\em Journal of Computer and System Sciences}}

\end{document}